\documentclass[fleqn,usenatbib]{mnras}
\usepackage{newtxtext}
\usepackage{mathptmx}
\usepackage{mathtools}
\usepackage{txfonts}
\usepackage[T1]{fontenc}
\usepackage{ae,aecompl}
\usepackage{xcolor}

\DeclareRobustCommand{\VAN}[3]{#2}
\let\VANthebibliography\thebibliography
\def\thebibliography{\DeclareRobustCommand{\VAN}[3]{##3}\VANthebibliography}

\usepackage{graphicx}	
\usepackage{amsmath}	
\usepackage{amssymb}
\usepackage{bm}
\usepackage{pdflscape}

\title{An accurate equation for the gravitational bending
of light by a static massive object}

\author[O. del Barco]{Oscar del Barco\thanks{E-mail: obn@um.es}
\\
Laboratorio de \'{O}ptica, Instituto Universitario de Investigaci\'{o}n en \'{O}ptica y
Nanof\'{i}sica, Universidad de Murcia, Campus de Espinardo, E-30100, Murcia, Spain\\}

\pubyear{2024}

\begin{document}
\label{firstpage}
\pagerange{\pageref{firstpage}--\pageref{lastpage}}
\maketitle

\begin{abstract}
An exact analytical expression for the bending angle
of light due to a non-rotating massive object,
considering the actual distances from source and observer
to the gravitational mass, is derived.
Our novel formula generalizes Darwin's well-known
equation for gravitational light bending
[Proc. R. Soc. London A 263, 39-50 (1961)],
where both source and observer are placed
at infinite distance from the lensing mass,
and provides excellent results in comparison
with the post-Newtonian (PPN) formalism
up to first order. As a result, the discrepancy
between our recent expression and the PPN approach is
6.6 mas for sun-grazing beams coming from planet
Mercury, with significant differences up to 2 mas
for distant starlight. Our findings suggest that
these considerations should not be dismissed for both
solar system objects and extragalactic sources,
where non-negligible errors might be present
in ultraprecise astrometry calculations.
\end{abstract}


\section{Introduction}

The gravitational deflection of light by a massive
body has been a subject of intense research for
over three centuries. According to \citet{Newton1730},
if a light ray from a distant star passes near a massive body
it would be bent a very small amount due to the object’s
gravity. However, it was not until 1804 when this bending
angle was first calculated by \citet{Soldner1804}
resulting in a value of 0.87 arcsec for sun-grazing
starlight. A century later, \citet{Einstein1916} reported a deflection
angle of 1.75 arcsec within the framework of
the general theory of relativity (GR),
twice the value as obtained by Newtonian
mechanics. This result was experimentally
confirmed by Eddington from the May
1919 solar eclipse expeditions \citep{Eddington1920,Will2015}
and subsequent measurements via Very Long Baseline
Interferometry (VLBI), a technique capable
of measuring bending angles from distant
radio sources with high accuracy
\citep{Shapiro1967,Lebach1995,Li2022a,Li2022b}.

Apart from providing a means to test GR,
gravitational deflection of light has been widely employed to
observe the properties of very distant galaxies, as well as
to infer the mass of astrophysical objects, since the
massive body acts as a gravitational lens with a characteristic
magnifying effect \citep{Schneider1992,Frittelli2003,Ye2008}.
Accordingly, gravitational lensing is indeed a milestone
of astronomy with wide-ranging applications, covering extra-solar
planets \citep{Turyshev2022}, black hole lensing
\citep{Iyer2007,Virbhadra2009}
or string theory \citep{He2022,Kong2024}.
In order to calculate the light deflection angle for
static massive objects, the prevalent GR formalism considers
that the path of a light ray is a null geodesic in
different manifolds
\citep{Misner1973,Chandrasekhar1983,Wald1984,Bozza2005}.

An alternative method to study the effect
of gravity on light is the so-called
material medium approach (hereafter MMA),
based on the idea of representing the gravitational
field as an optical medium with an effective refractive
index. In fact, this different conception
of light bending has a long history since
the early days of general relativity
\citep{Eddington1920,Whitehead1922}. Eddington himself
admitted that the gravitational deflection
effect on light could be imitated by a
refractive medium filling the space round
the Sun, giving an appropriate velocity
of light. Specifically, this refractive
index at a distance $r$ from the center
of the Sun should be $[1-(r_{\rm s}/r)]^{-1}$,
where $r_{\rm s}$ corresponds to the Sun's
Schwarzschild radius \citep{Eddington1920}. Therefore,
a light ray passing through a material
medium will be deviated due to the refractive
index variation of the associated media,
in accordance with the well-established
general relativity explanation.

Apart from the Eddington's analysis on
gravitational light bending, the MMA
was first developed by Tamm during
the 1920s \citep{Tamm1924,Tamm1925}. This innovative
idea was used by several authors to discuss
the optical phenomena for the deflection of
electromagnetic waves by a gravitational
field \citep{Balazs1958,Plebanski1960,deFelice1971}, mainly for
non-rotating masses in the Schwarschild geometry
\citep{Fischbach1980,Nandi1995,Evans1996a,Evans1996b,Sen2010,Feng2019,Feng2020},
where the medium refractive index could be
expressed as an infinite power series of
$(r_{\rm s}/r)$ terms
\citep{Schneider1992,Petters2002,Roy2019,Meneghetti2021,Hwang2024}.
Moreover, the same method was also applied to
estimate the light deflection angle caused by
the rotation of gravitating bodies \citep{Roy2015}
or charged massive objects \citep{Roy2017}.

In both theoretical frameworks (i.e., the material
medium approach and general relativity) some authors
consider an asymptotic scenario where source and
observer are placed at infinite distance from the
lensing mass, which is actually a reasonable approach given
the large distances involved. On the other hand,
some researchers have studied a finite-distance
scheme with specific light paths, addressing the
problem numerically \citep{Feng2019,Feng2020} or via approximate
deflection angle equations
\citep{Zschocke2011,Ishihara2017,Takizawa2023}.
In this regard, the formalism provides a means
to analytically calculate the bending angle to any
desired order of accuracy by expanding
a general formula in $M/r$ terms \citep{Cowling1984},
where $M$ is the mass of the gravitational body.
In particular, an approximate first order PPN equation
has been widely used throughout the literature to determine
precise bending angle calculations
\citep{Will1993,Ni2017,Li2022a,Li2022b}.
Nevertheless, to the best of our knowledge,
an exact analytical expression for the light
deflection angle due to a stationary massive body,
within a finite-distance scenario, has not yet
been reported.

In this article, we employ the MMA formalism to
derive an accurate equation for the gravitational
deflection of light by a static massive object
which generalizes Darwin's well-known formula
\citep{Darwin1961}, where infinite distances
from source and observer to the gravitational
mass are assumed. As a result, we show that
non-negligible errors in the positioning of
celestial objects should be avoided if we
take into consideration our recent equation.
Furthermore, we also test the validity of
our material medium approach via numerical
calculation of the gravitational time delay
of light (commonly named the Shapiro time delay).

The paper is organized as follows. In section 2
we present our MMA method to deduce an exact
analytical equation for the gravitational
deflection angle of light in the Schwarschild
spacetime, considering the actual distances from
source and observer to the gravitational body.
Moreover, the Shapiro time delay is also revisited
within the MMA framework. In section 3 we
describe our fundamental analytical and numerical
results, where the appropriateness of our novel
equation in ultraprecise astrometry is highlighted.
Finally, we summarize our main results and
conclusions in section 4.

\section{Theoretical basis}

In this section we develop our material medium
approach to derive a general equation
for the bending angle of light due to a static
massive object. Then, the Shapiro time delay is
addressed theoretically for completeness.
\begin{figure*}
\begin{center}
\includegraphics[width=.75\textwidth]{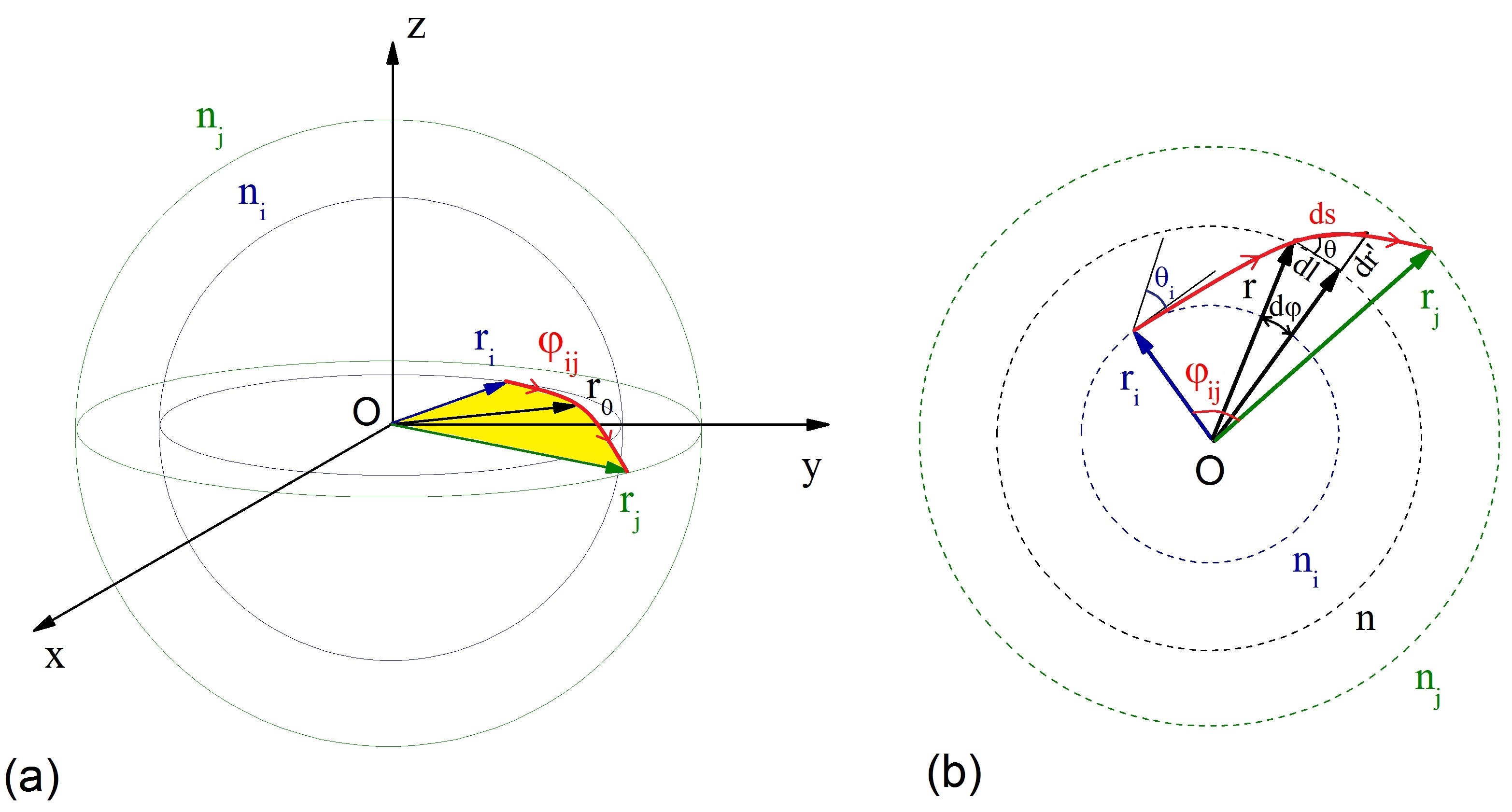}
\caption{(a) Schematic representation of our
spherical-stratified medium in a non-flat
spacetime (i.e., the Schwarschild spacetime
in our case). The gravitational mass $M$
is located at point O and each concentric
sphere presents the same refractive index
value, according to equation~(\ref{refin}).
The light source and observer are placed at
coordinates $r_i$ and $r_j$, respectively, and a light
ray propagates between both positions. The closest
approach distance to the static mass is $r_0$, whereas
the rotation angle is given by $\varphi_{ij}$.
Without loss of generality, we restrict ourselves
to the equatorial plane where the polar angle is
$\pi/2$ (b) top view of the equatorial plane
with the elements required to compute
the propagation trajectory of a light
ray in this graded medium.}\label{fig1}
\end{center}
\end{figure*}

\subsection{Material medium approach}

Let us first analyze Fig.~\ref{fig1}(a)
a spherical-stratified medium in a non-flat
spacetime (i.e., the Schwarschild spacetime
in our case). The gravitational
mass $M$ is located at point O and modifies the
refractive index of the surrounding media,
in accordance with the MMA.
For our static body, this index of refraction
$n(r)$ is spherically symmetric and depends
exclusively on the radial coordinate $r$
and its Schwarschild radius $r_{\rm s}=2GM/c^2$,
where $G$ is the gravitational constant and
$c$ the speed of light in vacuum. In this context,
a light ray describes a specific path in this
graded medium where $r_i$ and $r_j$ indicate
the positions of source and observer, respectively.
The parameter $r_0$ is the closest approach
distance to our lensing mass, whereas
$\varphi_{ij}$ stands for the angle between both
locations. For the sake of simplicity,
we restrict ourselves to the equatorial
plane where the polar angle is $\pi/2$.

A detailed description of this plane,
with the elements required to compute
the propagation trajectory of a light
ray in a graded medium, is depicted in
Fig.~\ref{fig1}(b). First, we need
to pay special attention to the
relation between the radial length
of the light's infinitesimals \citep{Misner1973}
\begin{equation}\label{GO1}
\mathrm{d}r'=\mathrm{d}r \left(1-
\frac{r_{\rm s}}{r}\right)^{-1/2} =
\mathrm{d}s \sin\theta,
\end{equation}
where $\mathrm{d}r'$ corresponds to the
proper distance in our curved spacetime,
unlike the coordinate distance
$\mathrm{d}r$ (not shown in this
figure) applicable to a flat space.
Moreover, $\mathrm{d}s$ denotes the
length of an infinitesimal ray path
and $\mathrm{d}l$ is the lateral
length.

Considering the basic relation
between the infinitesimal angle
$\mathrm{d}\varphi$ and $\mathrm{d}s$
\citep{Feng2019}
\begin{equation}\label{GO2}
r \mathrm{d}\varphi=\mathrm{d}s \cos\theta,
\end{equation}
and the previous equation~(\ref{GO1}),
we trivially obtain the following
differential equation for light
propagation
\begin{equation}\label{GO3}
r \frac{\mathrm{d}\varphi}{\mathrm{d}r}=
\left(1-\frac{r_{\rm s}}{r}\right)^{-1/2}
\cot\theta,
\end{equation}
which satisfies the renowned Bouguer's law
in geometric optics
\begin{equation}\label{GO4}
n_i r_i \cos\theta_i = n r \cos\theta = q.
\end{equation}
Here, the impact parameter of the light ray
$q$ is a constant for a spherical-stratified
material medium $n=n(r)$, as in our case.
Accordingly, we can express equation~(\ref{GO3}) as
\begin{equation}\label{DEGO}
\frac{\mathrm{d}\varphi}{\mathrm{d}r} =
\frac{q}{r} \frac{1}{\sqrt{n^2 r^2-q^2}}
\left(1-\frac{r_{\rm s}}{r}\right)^{-1/2}.
\end{equation}

Let us now assume that the medium refractive
index $n(r)$ is just the positive square root of
Eddington's proposal \citep{Eddington1920}
\begin{equation}\label{refin}
n(r)=\left(1-\frac{r_{\rm s}}{r}\right)^{-1/2}.
\end{equation}
Introducing equation~(\ref{refin}) into
equation~(\ref{DEGO}), the general differential
equation for light propagation in such
a stratified medium can be written in the
following way
\begin{equation}\label{DEGOdef}
\frac{\mathrm{d}\varphi}{\mathrm{d}r} = \frac{1}{r^2}
\left[\frac{r^3-q^2(r-r_{\rm s})}{q^2
r^2(r-r_{\rm s})}\right]^{-1/2}
\left(1-\frac{r_{\rm s}}{r}\right)^{-1/2},
\end{equation}
which depends on the mass of our
central object and the impact parameter
of the light ray.

Within the scope of GR theory,
a light beam in a Schwarschild
spacetime obeys the following ordinary
differential equation in the equatorial
plane \citep{Misner1973}
\begin{equation}\label{DEGR}
\frac{\mathrm{d}\varphi}{\mathrm{d}r} = \frac{1}{r^2}
\left[\frac{r^3-q^2(r-r_{\rm s})}{q^2 r^3}\right]^{-1/2}
\left(1-\frac{r_{\rm s}}{r}\right)^{-1/2},
\end{equation}
where, again, $q$ corresponds to the impact
parameter of the light ray. Please, note the
similarity between equations~(\ref{DEGOdef}) and
(\ref{DEGR}). In fact, a complete equivalence
is achieved for the weak-field approximation
when $r>>r_{\rm s}$. In other words, our MMA formalism
reproduces exactly the propagation of light
in a gravitational field, provided
that the ray paths are sufficiently distant from the
Schwarschild radius of the non-rotating body.
This requisite is fulfilled in our astronomical
scenarios provided that $r_{\rm s} = 2.9$ km for the Sun
and $r=695700$ km for sun-grazing light beams.

Hence, the angle $\varphi_{ij}$ between positions
$r_i$ and $r_j$ can be accurately evaluated via
the material medium approach \citep{Feng2019}
(please, see again Fig.~\ref{fig1}(b))
\begin{equation}\label{angleij}
\varphi_{ij}=\int_{r_i}^{r_j} \mathrm{d}\varphi=
\int_{r_i}^{r_j} \frac{\mathrm{d}l}{r}=
\int_{r_i}^{r_j} \frac{\mathrm{d}r'}{r \tan\theta}.
\end{equation}
Substituting Bouguer's law, equation~(\ref{GO4}), into
equation~(\ref{angleij}) and performing
some elementary algebra, we obtain
\begin{equation}\label{angleijdef}
\varphi_{ij}=\int_{r_i}^{r_j} \mathrm{d}r
\frac{n_0r_0}{r \sqrt{n^2r^2-n_0^2r_0^2}}
\left(1-\frac{r_{\rm s}}{r}\right)^{-1/2},
\end{equation}
where the refractive index $n(r)$ is given
by equation~(\ref{refin}). It is worth mentioning
that equation~(\ref{angleijdef}) has analytical
solutions in terms of incomplete elliptic
integrals of first kind, as briefly addressed.

\begin{figure}
\begin{center}
\includegraphics[width=.48\textwidth]{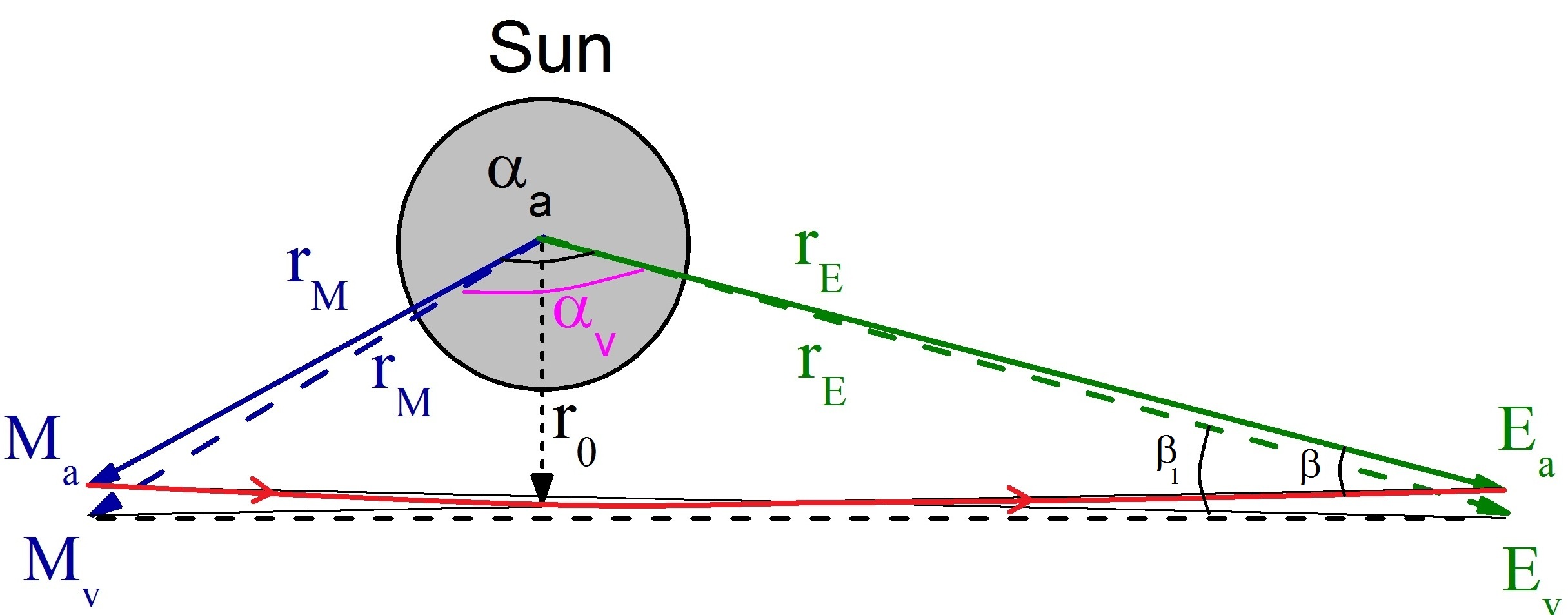}
\caption{Our MMA method applied to a suitable scenario
in the solar system. A light ray coming from Mercury
($\textrm{M}_{\rm a}$) passes close to the Sun and
reaches the Earth ($\textrm{E}_{\rm a}$). Due to the
gravitational light bending, a virtual position of each planet
occurs, represented by $\textrm{M}_{\rm v}$ and
$\textrm{E}_{\rm v}$. In this situation, the deflection
angle $\Delta\alpha^{\rm{(MMA)}}$ is computed as the
difference between the actual angle $\alpha_{\rm a}$ and
the virtual angle $\alpha_{\rm v}$ via
equations~(\ref{angleijdef}) and (\ref{GDGO}).}\label{fig2}
\end{center}
\end{figure}

Once the basic formalism has been introduced,
let us now study a specific example where
our new approach should be appropriate.
Our light source will be planet Mercury near its solar
superior conjunction, and a light ray
coming from this planet passes close to
the Sun and reaches the Earth.
This situation is illustrated in
Fig.~\ref{fig2} where our light
beam travels from the actual position of Mercury
($\textrm{M}_{\rm a}$) to the Earth
($\textrm{E}_{\rm a}$). It should be remarked
that, due to the gravitational light bending,
an observer on Earth would experience a
virtual position of Mercury ($\textrm{M}_{\rm v}$).
The inverse scenario is fully applicable,
where now $\textrm{E}_{\rm v}$ stands for
the virtual location of the Earth.
Furthermore, the average distances from the Sun
to each planet are denoted by $r_{\rm M}$
and $r_{\rm E}$, $r_0$ stands for the closest approach
of the ray path to the Sun, and $\beta$
is the angle between Mercury
(in the absence of gravitational bending)
and the Sun as seen by an Earth's
observer \citep{Li2022a,Li2022b}.

Therefore, the deflection angle
$\Delta\alpha^{\rm{(MMA)}}$
is calculated as the difference between
the actual angle $\alpha_{\rm a}$ and
the virtual angle $\alpha_{\rm v}$
\citep{Feng2019}
\begin{eqnarray}\label{GDGO}
\Delta\alpha^{\rm{(MMA)}}&=& \alpha_{\rm a}-\alpha_{\rm v} =
(\varphi_{0{\rm M}}+\varphi_{0{\rm E}})
\nonumber\\
&-&\left[\arccos\left(\frac{r_0}{r_{\rm M}}\right)+
\arccos\left(\frac{r_0}{r_{\rm E}}\right)\right],
\end{eqnarray}
where $\varphi_{0{\rm M}}$ and $\varphi_{0{\rm E}}$
correspond to the angle between $r_0$ and each
planet's actual position, evaluated via
equation~(\ref{angleijdef}). As the difference
between the angles $\alpha_{\rm a}$ and $\alpha_{\rm v}$
is very small, we can consider that
$\beta \simeq \beta_{1}$ in order to
determine a practical expression for the
observation angle $\beta$.

Additionally, the deflection angle under the PPN formalism
up to first order reads \citep{Li2022a,Li2022b}
\begin{equation}\label{GDPPM}
\Delta\alpha^{\rm{(PPN)}}= (1+\gamma) \frac{G M_{\rm s}}{r_0 c^2}
\left(\cos\beta - \cos\delta\right),
\end{equation}
where $M_{\rm s}$ is the solar mass, $\sin\beta=r_0/r_{\rm E}$,
$\sin\delta=r_0/r_{\rm M}$, and $\gamma$ stands for
the dimensionless PPN parameter used to characterize
the contribution of the spacetime curvature to
the gravitational deflection. In this regard,
we will assume that $\gamma=1$ (as theoretically
established in GR) since this choice
does not influence significatively
the results of the positions of celestial bodies
in the solar system \citep{Li2022a}.

When $r_0$ is far less than the distance
from the Sun to both the Earth (observer)
and Mercury (source), equation~(\ref{GDPPM})
transforms into the celebrated Einstein's
formula up to first order
\citep{Einstein1916,Misner1973,Mutka2002}
\begin{equation}\label{GDEins}
\Delta\alpha^{\rm{(Ein)}}=\frac{2r_{\rm s}}{r_0},
\end{equation}
where a deflection angle of 1.7518 arcsec for starlight
grazing the solar surface has been universally accepted.
As shortly discussed, a detailed comparison between
the first order PPN formula, equation~(\ref{GDPPM}),
and our exact MMA expression will be carried out.

\subsection{Exact analytical equation for the
bending angle}

The analytical solution of equation~(\ref{angleijdef}) has
been obtained via a Wolfram Mathematica software.
After a careful analysis, the physically acceptable
solution for the angle $\varphi_{ij}$ is given by
\begin{eqnarray}\label{angleijanly}
\varphi_{ij}&=&2 n(r_i) \sqrt{\frac{r_0(r_i-r_{\rm s})}
{r_i Q}} \ F(z_i,k)
\nonumber\\
&-& 2 n(r_j) \sqrt{\frac{r_0(r_j-r_{\rm s})}
{r_j Q}} \ F(z_j,k),
\nonumber\\
\end{eqnarray}
where $Q^2=(r_0-r_{\rm s})(r_0+3r_{\rm s})$ and $F(z,k)$
is the Legendre elliptic integral of the first kind,
with the following relations for the Jacobi
amplitude $z(r)$
\begin{equation}\label{Fz}
\sin^2 z = \frac{2r_0 r_{\rm s}+r(r_{\rm s}-r_0+ Q)}
{r(3r_{\rm s}-r_0+Q)},
\end{equation}
and the elliptic modulus $k$
\begin{equation}\label{Fm}
k^2=\frac{3r_{\rm s}-r_0+Q}{2Q}.
\end{equation}

So, equation~(\ref{GDGO}) can be rewritten as
(please, see again Fig.~\ref{fig2})
\begin{eqnarray}\label{GDfin}
\Delta\alpha^{\rm{(MMA)}}&=&\left[4 n_0 \sqrt{\frac{r_0-r_{\rm s}}
{Q}} \ F\left(\frac{\pi}{2},k\right) \right.
\nonumber\\
&-& \left. 2 n(r_{\rm M}) \sqrt{\frac{r_0(r_{\rm M}-r_{\rm s})}
{r_{\rm M} Q}} \ F(z(r_{\rm M}),k) \right.
\nonumber\\
&-& \left. 2 n(r_{\rm E}) \sqrt{\frac{r_0(r_{\rm E}-r_{\rm s})}
{r_{\rm E} Q}} \ F(z(r_{\rm E}),k)\right]
\nonumber\\
&-& \left[\arccos\left(\frac{r_0}{r_{\rm M}}\right)+
\arccos\left(\frac{r_0}{r_{\rm E}}\right)\right].
\end{eqnarray}
In the asymptotic case, that is, when both
source and observer are placed at infinite
distance from the gravitational body,
equation~(\ref{GDfin}) reduces to the well-known
Darwin's formula \citep{Darwin1961,Misner1973,Mutka2002}
\begin{equation}\label{GDDarwin}
\Delta\alpha^{\rm{(Dar)}}=4 \sqrt{\frac{r_0}{Q}}
\left[F\left(\frac{\pi}{2},k\right)-
F(z_{\infty},k)\right]-\pi,
\end{equation}
by just considering $r_{\rm E}, r_{\rm M}
\to \infty$ in our generalized formula,
where now
\begin{equation}\label{zinf}
\sin z_{\infty}^2=\frac{r_{\rm s}-r_0+ Q}
{3r_{\rm s}-r_0+Q}.
\end{equation}

If only the light source distance to the
lensing mass is significantly higher than
the closest approach $r_0$, we can take
the limit $r_{\rm M} \to \infty$ to obtain
a modified version of equation~(\ref{GDfin})
\begin{eqnarray}\label{GDfinin}
\Delta\alpha^{\rm{(MMA)}}&=&\left[4 n_0 \sqrt{\frac{r_0-r_{\rm s}}
{Q}} \ F\left(\frac{\pi}{2},k\right) \right.
\nonumber\\
&-& \left. 2 n(r_{\rm E}) \sqrt{\frac{r_0(r_{\rm E}-r_{\rm s})}
{r_{\rm E} Q}} \ F(z(r_{\rm E}),k) \right.
\nonumber\\
&-& \left. 2 \sqrt{\frac{r_0} {Q}} \ F(z_{\infty},k)\right]-
\left[\arccos\left(\frac{r_0}{r_{\rm E}}\right)+\frac{\pi}{2}\right],
\end{eqnarray}
which constitutes an excellent tool to accurately
calculate the gravitational deflection angle
for extragalactic emitters.
\begin{figure}
\begin{center}
\includegraphics[width=.48\textwidth]{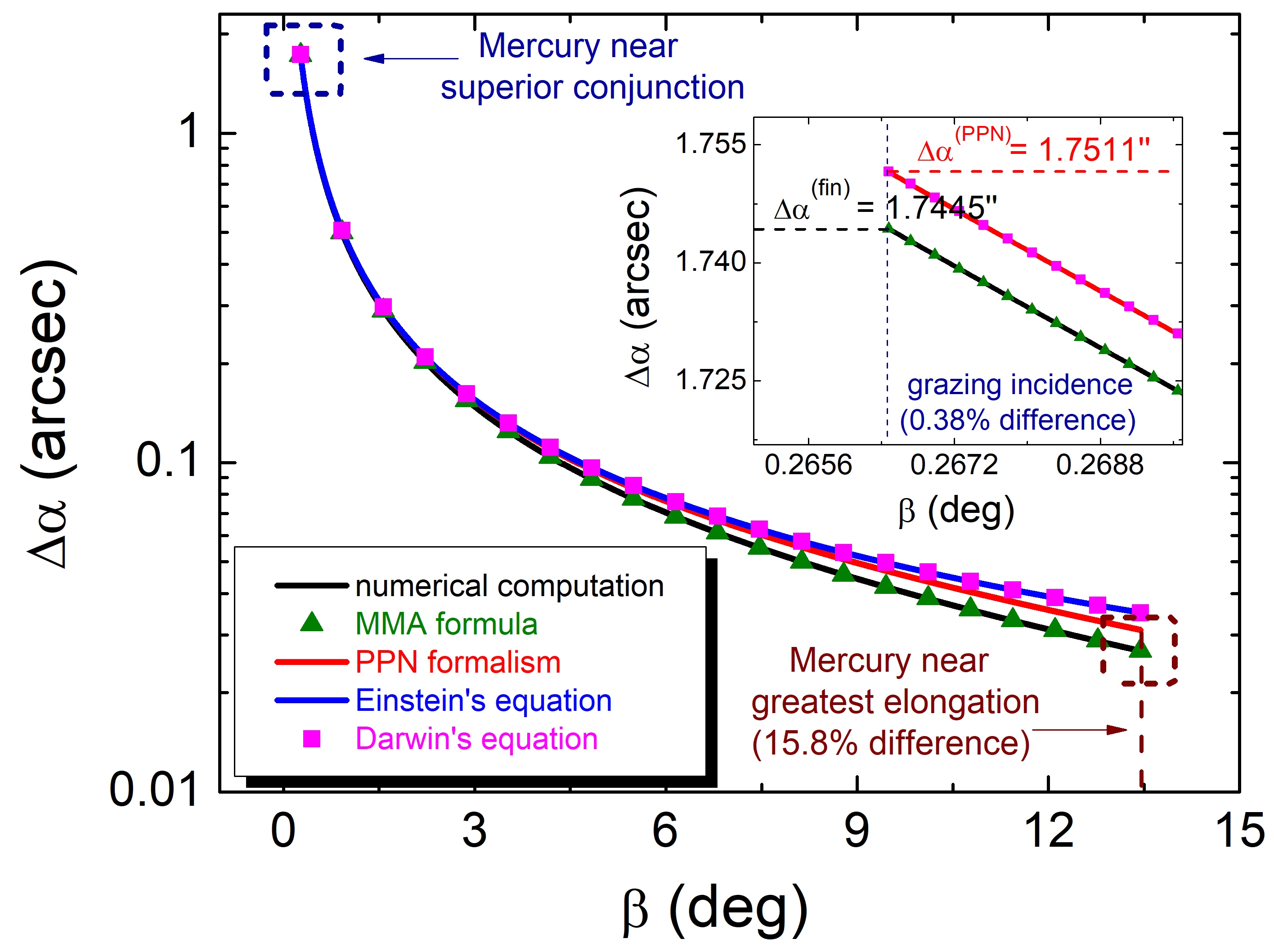}
\caption{The gravitational deflection angle
$\Delta\alpha$ for a light beam coming from
Mercury, passing near the Sun and reaching
the Earth, versus the observation angle $\beta$
(please, see again Fig.~\ref{fig2}). The numerical
computation of equation~(\ref{GDGO})
and the PPN formula up to first order, equation~(\ref{GDPPM}),
differ for higher values of $\beta$. Moreover, the MMA
equation~(\ref{GDfin}) based on elliptical
integrals fits precisely our numerical
calculations. For completeness,
the deflection angle results derived
via Darwin's formula, equation~(\ref{GDDarwin}),
are also depicted. The inset shows
the situation for a sun-grazing beam,
where now the error between both schemes
reduces to 0.38\%.}
\label{fig3}
\end{center}
\end{figure}

\subsection{Shapiro time delay calculation}

So far, we have analyzed the gravitational
bending angle on the basis of our
MMA method. Let us now investigate another
crucial parameter related to the effect of
gravitational bodies on light propagation.
We are referring to the Shapiro time delay
$\Delta t$, the relativistic
time shift in the round-trip travel time
for light signals reflecting off other planets
\citep{Shapiro1964,Shapiro1971,Reasenberg1979}.
According to the astronomical scenario described in
Fig.~\ref{fig2}, $\Delta t$ can be easily
expressed as
\begin{equation}\label{Shaptimed}
\Delta t = 2\left[(t_{0{\rm M}}+t_{0{\rm E}})-
\frac{1}{c} \left(\sqrt{r_{\rm M}^2-r_0^2}+
\sqrt{r_{\rm E}^2-r_0^2}\right)\right],
\end{equation}
where $t_{0{\rm M}}$ and $t_{0{\rm E}}$
stand for the light propagation time
between $r_0$ and each planet's
actual position (i.e., Mercury and
the Earth in our situation).
\begin{figure*}
\begin{center}
\includegraphics[width=.99\textwidth]{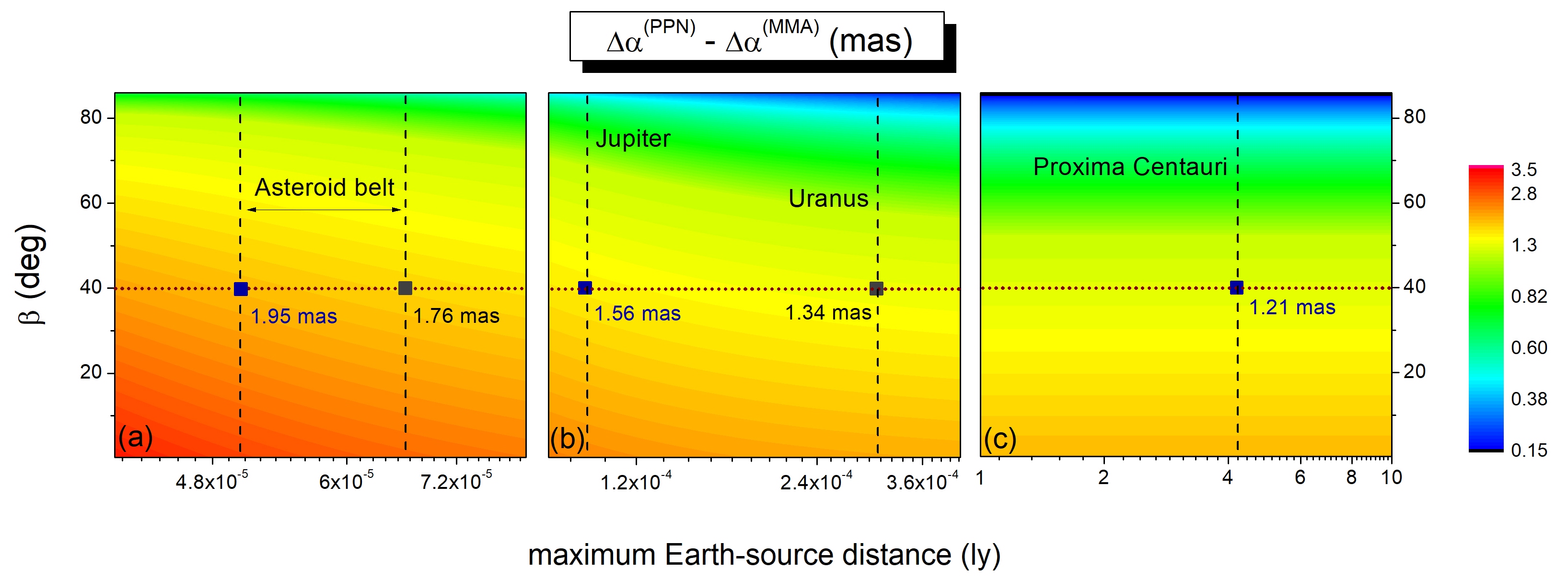}
\caption{Absolute difference between the first order
PPN formalism and our MMA method for $\Delta\alpha$
in the case of solar system objects (left and central
panels) and extrasolar sources (right panel).
This parameter has been computed via
equations~(\ref{GDPPM}) and (\ref{GDfin}) for
Figs.~\ref{fig4}(a) and Fig.~\ref{fig4}(b),
whereas our asymptotic formula, equation~(\ref{GDfinin}),
has been used for distant starlight calculations
(right panel) as a function of the angle $\beta$ and
the maximum source-Earth distance in light years.
It can be noticed that the contours
in the right panel are flat due to the large distances
involved (that is, the source is infinitely far away
in all cases), but our results are non-zero and depend
on the angle $\beta$ because the Earth is not at infinite distance from the
Sun, compared to the impact parameter of the light ray.}\label{fig4}
\end{center}
\end{figure*}

The usual way to deduce an exact
expression for the time $t_{ij}$
that a light ray takes to travel from
from position $r_i$ to $r_j$
is through general relativity considerations
\citep{Wald1984}
\begin{equation}\label{ShapGR}
t_{ij}^{(\rm{GR})}=\frac{\sqrt{r_j^2-r_i^2}}{c}+
\frac{r_{\rm s}}{c} \log\left(\frac{r_j+
\sqrt{r_j^2-r_i^2}}{r_i}\right)
+ \frac{r_{\rm s}}{2c} \
\sqrt{\frac{r_j-r_i}{r_j+r_i}},
\end{equation}
nevertheless, we can also apply our
MMA formalism to compute
these time lapses in an alternative
manner.

Looking back at Fig.~\ref{fig1}(b),
the parameter $t_{ij}$ can be evaluated
as \citep{Feng2019,Feng2020}
\begin{equation}\label{tij}
t_{ij}^{(\rm{MMA})}=\int_{r_i}^{r_j} \frac{n
\mathrm{d}s}{c}= \int_{r_i}^{r_j}
\frac{n \mathrm{d}r'}
{c\sqrt{1-\cos^2\theta}},
\end{equation}
which transforms into the following
expression, once the relation between
$\mathrm{d}r'$ and $\mathrm{d}r$,
equation~(\ref{GO1}), has been
considered
\begin{equation}\label{tijdef}
t_{ij}^{(\rm{MMA})}=\frac{1}{c}
\int_{r_i}^{r_j} \mathrm{d}r
\frac{n^{2} r}{\sqrt{n^2r^2-n_0^2r_0^2}}
\left(1-\frac{r_{\rm s}}{r}\right)^{-1/2},
\end{equation}
where, again, the refractive index $n(r)$ is
given by equation~(\ref{refin}).

As in the case of the gravitational
light bending, equation~(\ref{tijdef})
has analytical solutions in terms of
incomplete elliptic integrals of different
kinds. However, due to the mathematical
complexity of the final expression
and its limited usefulness, we have
not included this new equation in
our article. As briefly discussed,
we will show the equivalence between
the GR formula for the Shapiro
time delay, equation~(\ref{ShapGR}),
and our numerical calculations via
equation~(\ref{tijdef}).

\section{Results}

In this section we present some analytical
and numerical results concerning the
gravitational light bending and Shapiro
time delay via our MMA formalism. As a
consequence, we want to emphasize
the advisability of using our new
analytical expressions to prevent
undesired ultraprecise
astrometry errors.

Hence, we represent in Fig.~\ref{fig3}
the gravitational deflection angle
for light beams coming from Mercury
(please, see again Fig.~\ref{fig2}),
where $\Delta\alpha$ is shown as a function
of the angle $\beta$. One notices that the
numerical computation of equation~(\ref{GDGO})
(black solid curve) and Einstein's first
order formula equation~(\ref{GDEins})
(blue solid curve) differ substantially
for higher values of $\beta$. This discrepancy
is reduced in the case of the first order
PPN formalism, equation~(\ref{GDPPM}), where
the difference between both methods reaches
15.8\% when Mercury is located near its
greatest elongation. For completeness,
the deflection angle results derived
via Darwin's formula, equation~(\ref{GDDarwin}),
are also depicted in Fig.~\ref{fig3},
where a full agreement with Einstein's
equation is attained.

It can be noticed that our MMA formula,
equation~(\ref{GDfin}), fits precisely to the
numerical calculation of the deflection
angle equation~(\ref{GDGO}). Moreover,
when Mercury is at its superior
conjunction (i.e., $\beta \simeq 0.26$ deg),
the difference between both methods also
exists but to a lesser extent, as appreciated
in the figure inset. Accordingly, a discrepancy
of 0.38\% is achieved for solar grazing incidence
if we do not consider our MMA equation.
Our previous results indicate the
importance of taking into consideration
our exact analytical formula when
calculating the gravitational deflection
angle $\Delta\alpha$, especially when
solar system distances are involved.
In fact, non-negligible errors are also
presented when extrasolar distances are
considered, as explained ahead.
\begin{figure}
\begin{center}
\includegraphics[width=.46\textwidth]{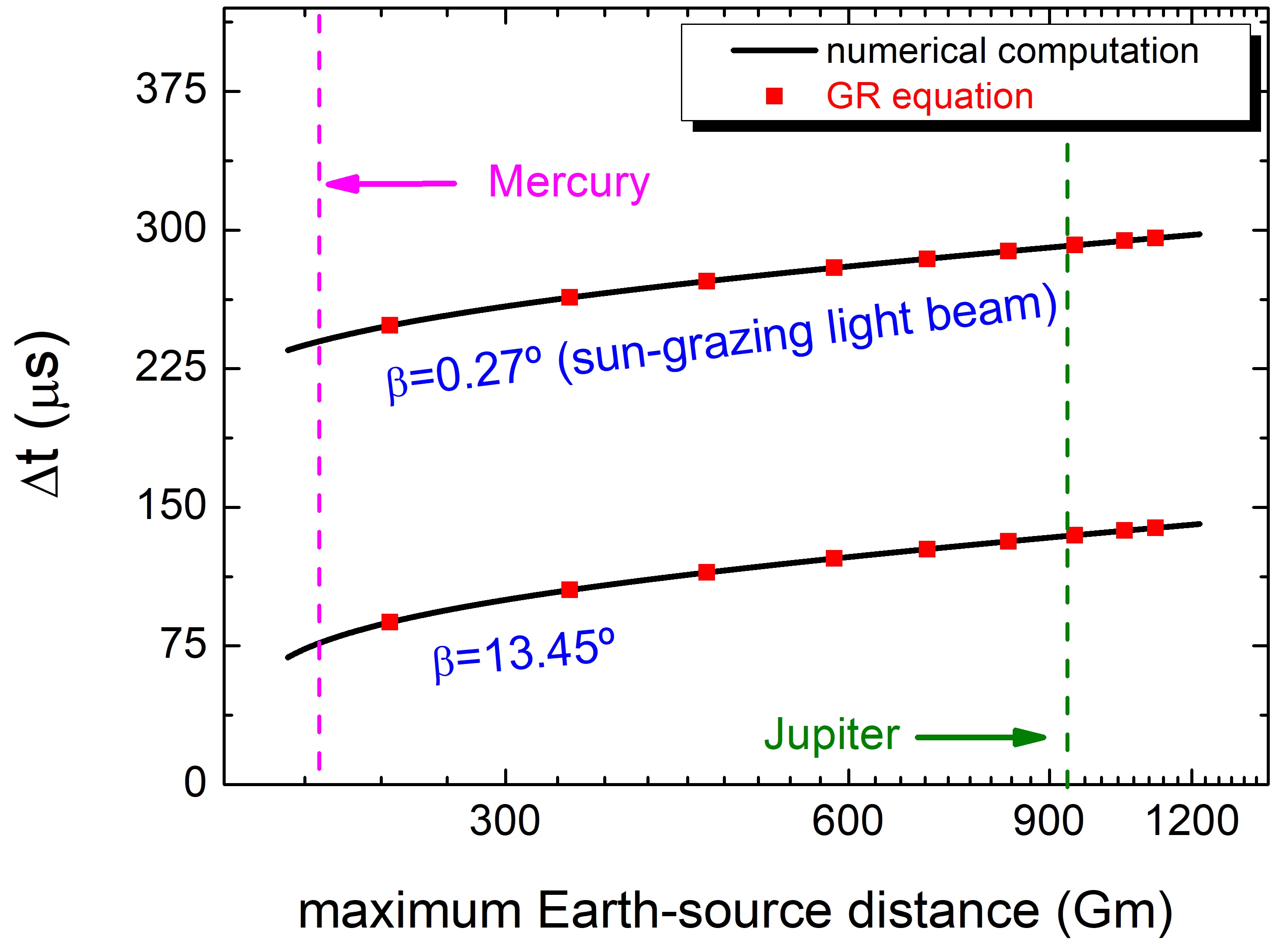}
\caption{Shapiro time delay $\Delta t$ versus
the maximum source-Earth distance evaluated
numerically via our MMA model,
equations~(\ref{Shaptimed}) and (\ref{tijdef}),
for two different observation angles $\beta$.
The red squares indicate the $\Delta t$ values
computed via GR formalism,
equations~(\ref{Shaptimed}) and (\ref{ShapGR}).
A full agreement between both theoretical
models is observed.}\label{fig5}
\end{center}
\end{figure}

Consequently, we have studied in Fig.~\ref{fig4}
the absolute difference between the first order
PPN formalism and our MMA method for
$\Delta\alpha$ in the case of solar system objects
(left and central panels) and extrasolar sources
(right panel). This parameter has been computed via
equations~(\ref{GDPPM}) and (\ref{GDfin}) for
Figs.~\ref{fig4}(a) and \ref{fig4}(b),
whereas our asymptotic formula, equation~(\ref{GDfinin}),
has been used for distant starlight calculations
(right panel) as a function of the angle $\beta$ and
the maximum source-Earth distance in light years (ly).
For an observation angle of 40 deg, a discrepancy
between 1.76 and 1.95 mas is attained for solar
system sources within the asteroid belt distance,
while a lower deviation of 1.21 mas is also
encountered for Proxima Centauri, as shown in
Fig.~\ref{fig4}(c). Furthermore, significant
differences are also reported for sun-grazing
light beams coming from solar system planets
like Jupiter (2.35 mas) and Uranus (2.11 mas),
dropping to 0.20 mas for extrasolar light
emitters at larger observation angles
($\beta \simeq 80$ deg).

On the other hand, the Shapiro time delay
$\Delta t$ for different source-Earth distances
is depicted in Fig.~\ref{fig5}, where the
concrete examples of Mercury and Jupiter
are illustrated. As in Fig.~\ref{fig4},
we have assumed the maximum distances
from emitter to observer. The solid lines
represent our MMA results performed numerically via
equations~(\ref{Shaptimed}) and (\ref{tijdef}),
whereas the squares indicate the $\Delta t$
values computed via general relativity formalism,
equations~(\ref{Shaptimed}) and (\ref{ShapGR}).
A total agreement between both models is observed
for different $\beta$ angles, noting the
appropriateness of our material medium
approach to describe light propagation in the
presence of static gravitational masses.

\section{Discussion and Conclusions}

Summarizing, a material medium approach
has been developed to determine an exact analytical
expression for the bending angle of light
due to a static massive body, considering
the actual distances from source and observer
to the gravitational mass. The validity of
our new method has been checked
throughout this article.

It is worth mentioning that a key conclusion
of our work is the desirability of taking into
account our novel accurate expressions,
equations~(\ref{GDfin}) and (\ref{GDfinin}),
when calculating the gravitational deflection angle
of light. In fact, relevant errors in the
positioning of celestial objects may occur
if our model is overlooked,
as presented in Figs.\ \ref{fig3} and \ref{fig4}.
For instance, the absolute difference between the
MMA method and the first order PPN formalism
at an observation angle of 40 deg is 1.21 mas
for starlight coming from Proxima Centauri,
while the angular diameter of this star is
about 1 mas \citep{Segransan2003}.
In this respect, a precise location
of this star might help to accurately estimate its
wide binary orbit around $\alpha$ Centauri A and B
\citep{Banik2019}.

Moreover, this bending angle inaccuracy is also
greater than $\Delta\alpha$ disagreement when
modelling our gravitational mass as a static
or a rotating body. Indeed, as reported by
Roy and Sen within the framework of an
asymptotic-based MMA in Kerr geometry
\citep{Roy2015}, the deflection angle for
distant starlight grazing the Sun is 1.7520
arcsec for light ray prograde orbits,
whereas 1.7519 arcsec is achieved in a
retrograde scenario. Provided the bending
angle value of 1.7512 arcsec for a
stationary gravitational object via
the first order PPN formalism, the
corresponding deviation if one neglects solar
rotation is roughly 0.8 mas, in comparison with
an absolute difference of 2 mas when our
MMA equation is obviated.

Besides the assumption of a non-rotating central
mass, it should be stated that the principal constraint
of our MMA model comes from the aforementioned
weak-field approximation, that is, when $r>>r_{\rm s}$.
This means that our new approach cannot explain
the strong deflection of light by a central mass,
where the bending angles are not small
\citep{Bisnovatyi-Kogan2015}. In this situation,
light beams trajectories are relatively close
to $r_{\rm s}$ (as in the case of a Schwarschild
black hole) and several turns near the photon
sphere are completed before
reaching the observer. As a consequence,
$\Delta\alpha=2m\pi$ rad for an integer $m$,
a physical phenomenon beyond the scope of our work.

Despite all our calculations in this article
are based on light deflection by the Sun,
the gravitational light bending by massive
objects in the solar system, such as planet
Jupiter, has recently gained a great deal of attention
\citep{Crosta2006,Brown2021,Li2022a,Li2022b},
due to its potential applications in microarcosecond
astrometry. After a detailed comparison between
our MMA equation and the first order PPN formalism
for distant starlight grazing Jupiter's limb,
we conclude that the difference between both
methods is roughly 0.002 $\mu$as, far beyond
the milliarcosecond regime described in this article.
However, this discrepancy should be significant
in future sub-microarcosecond accuracy for the
gravitational bending of light \citep{Brown2013}.

It should be emphasized that the fundamental reason
for the difference between our MMA results and
previous theories is that the source and observer
are in general not infinitely far away, compared to
the impact parameter of the light ray at the deflecting
massive body, apart from the approximate character of
the PPN method discussed in this article.
In essence, our exact analytical expressions
might constitute useful tools to accurately
calculate the gravitational deflection angle
of light due to a static massive body, which
should be relevant to current and future
research in order to prevent undesired
errors in ultraprecise astrometry.

\section*{Acknowledgements}

The author would like to acknowledge
Miguel Ortu\~{n}o and Indranil Banik
for helpful discussions and assistance
with manuscript preparation. O. del Barco
also thanks research support
from Agencia Estatal de Investigaci\'{o}n
(PID2019-105684RB-I00, PID2020-113919RBI00),
Fundaci\'{o}n S\'{e}neca (19897/GERM/15) and
Departamento de Ciencia, Universidad
y Sociedad del Conocimiento del Gobierno de
Arag\'{o}n (research group $\textrm{E44-23R}$).

\section*{Data Availability}

Provided the theoretical nature of this paper,
all data and numerical results generated or analysed
during this study are included in this article
(and based on the references therein).
All the numerical calculations have been
carried out on the basis of a Fortran 90 compiler
and Mathematica software.

\bibliographystyle{mnras}

\bsp
\label{lastpage}
\end{document}